\documentclass[eqsecnum,preprint,amsmath,amssymb,nofootinbib]{revtex4}
\usepackage{slashed}

\begin{document}

\title{SIM(2) and SUSY}

\date{\today}

\author{Andrew G. Cohen} \email{cohen@bu.edu}
\affiliation{Physics Department\\  Boston University\\
  Boston, MA 02215, USA} \author{Daniel Z. Freedman}
\email{dzf@math.mit.edu} \affiliation{Department of Mathematics and
  Center for Theoretical Physics, Massachusetts Institute of
  Technology\\ Cambridge, MA 02139, USA}

\begin{abstract}
  The proposal of {\tt hep-ph/0601236}, that the laws of physics in
  flat spacetime need be invariant only under  a SIM(2) subgroup of the
  Lorentz group, is extended to include supersymmetry. $\mathcal{N}=1$
  SUSY gauge theories which include SIM(2) couplings for the fermions
  in chiral multiplets are formulated.  These theories contain two
  conserved supercharges rather than the usual four.
\end{abstract}

\preprint{MIT-CTP 3744}
\maketitle

\section{Introduction}
\label{sec:introduction}

One of the present authors (A.C.), together with S. L. Glashow, has
suggested \cite{Cohen:2006ky} that the laws of physics need not be
invariant under the full Lorentz group but rather under a SIM(2)
subgroup, whose Lie algebra contains the four generators $T_1=K_x +
J_y,\, T_2=K_y-J_x,\, J_z,$ and $K_z$, a situation they called ``Very
Special Relativity'' (VSR).  Terms in the Lagrangian that are
invariant only under this subgroup necessarily break discrete
symmetries, including CP.  Unlike most other subgroups, many
elementary aspects of special relativity, including particle
propagation, are preserved under SIM(2).  The theory embodies a new
mechanism for neutrino mass which conserves lepton number without
introducing additional sterile states. Application to the
$SU(2)_L\times U(1)$ gauge symmetry of the standard model may give the
electron an electric dipole moment detectable in future experiments,
and lead to observable effects in the end-of-spectrum behavior in
tritium $\beta$-decay. These ideas and their associated phenomenology
are discussed in \cite{Cohen:2006ky,Cohen:2006ir,prep}.

Our concern in this paper is the construction of supersymmetric field
theories that are translation and SIM(2) invariant, but not Lorentz
invariant.  We show that the spacetime symmetry of any conventional
${\cal N} =1$ SUSY gauge theory can be truncated to SIM(2) by
including the characteristic SIM(2) conserving but Lorentz violating
couplings for the fermions of chiral multiplets.  Such theories have
two conserved supercharges whose parameters are constant Majorana
spinors which satisfy $\gamma_\mu n^\mu \epsilon =0,$ where $n^\mu =
(1,0,0,1)$ specifies the null ray fixed by the SIM(2)
subgroup.\footnote {These theories are rather different from the
  theories with two supercharges proposed in
  \cite{Seiberg:2003yz}. Those theories require Euclidean signature,
  while our theories posses a null vector, and Lorentzian signature is
  essential.}

\section{SIM(2) Fermions and how they propagate}
\label{sec:sim2-fermions-how}

In this section we review the basic fermion wave equation of SIM(2)
symmetry. A free fermion is postulated in \cite{Cohen:2006ir} to obey
the equation of motion
\begin{equation}
  \label{eq:3}
  \left( \slashed{\partial} - m^2 \slashed{n}\frac{1}{n\cdot \partial}
    - M\right)\Psi(x)=0. 
\end{equation}
The term involving the fixed null vector $n^{\mu}\equiv(1,0,0,1)$
violates Lorentz invariance, along with T, P, CT and
CP-invariance. Since this null vector scales under a boost along the
$z$-axis, SIM(2) invariance requires that it appear homogeneously.
Applying the operator $\left( \slashed{\partial} - m^2
  \slashed{n}/(n\cdot \partial) + M\right)$ and using
$\{\slashed{n},\slashed{\partial}\} = 2 n\cdot \partial$ and
$\slashed{n}\slashed{n} =0$ leads to
\begin{equation}
  \label{eq:4}
(\square - 2m^2 -M^2)\Psi =0. 
\end{equation}
The fermion thus propagates as a massive particle with total square
mass $2m^2 + M^2$. Even when $M=0$, the plane wave solutions carry
time-like 4-momentum $p^\mu$, and the operator $1/(n\cdot \partial)$
is then well defined acting on physical wavefunctions.

There is no analogous SIM(2) modification for a free scalar field.
Thus a scalar super-partner $Z(x)$ of the fermion above should satisfy
\begin{equation}
  \label{eq:5}
  (\square - 2m^2 -M^2) Z = 0.
\end{equation}

\section{Basic SIM(2) SUSY}
\label{sec:basic-sim2-susy}

We consider a chiral multiplet with Majorana spinor field $\Psi(x)$,
(with chiral projections $L\Psi,~R\Psi$), complex scalar $Z(x)$, and
auxiliary field $F(x)$. The SIM(2) modification of standard SUSY is
extremely simple: the kinetic action of the free chiral multiplet is
the usual term $S_{\text{old}}$ plus a VSR modification
$S_{\text{new}}$:
\begin{subequations}
  \label{eq:6}
  \begin{gather}
    \label{eq:7}
    S = S_{\text{old}} + S_{\text{new}}\\  
    S_{\text{old}} = \int\! d^4\!x
    \left[
      -\partial^\mu \bar{Z}\partial_\mu Z
      +\bar{\Psi}\gamma^\mu L \partial_\mu\Psi +\bar{F}F
    \right]\\
    S_{\text{new}} = - m^2\int\! d^4\!x \left[ \bar{\Psi} \slashed{n}
      \frac{1}{n\cdot \partial}  L\Psi + 2 \bar{Z}Z\right].
  \end{gather}
\end{subequations}

This theory is invariant under the standard SUSY transformations
\begin{subequations}
  \label{eq:8}
  \begin{align}
    \delta Z &=  \sqrt{2}\bar{\epsilon} L \Psi &  \delta\bar{Z} &=
    \sqrt{2} \bar{\epsilon} R \Psi\\ 
    \label{eq:9}
    \delta L\Psi &= - \sqrt{2} L(\slashed{\partial} Z + F)\epsilon &
    \delta R\Psi &= -\sqrt{2} R(\slashed{\partial} \bar{Z} + \bar
    {F})\epsilon\\   
    \delta  F &= \sqrt{2} \bar{\epsilon}\, \slashed{\partial} L\Psi
    &     \delta\bar{F} &= \sqrt{2}\bar{\epsilon}\, \slashed{\partial}
    R\Psi
  \end{align}
\end{subequations}
provided that the spinor parameters satisfy
\begin{equation}
  \label{eq:10}
  \slashed{n} \epsilon = 0.  
\end{equation}
This condition decreases the number of conserved supercharges from
four to two. Consistency of this condition follows from
$\slashed{n}\slashed{n}=0$. 

To establish invariance we need only show that $\delta S_{\text{new}}$
vanishes under the transformation rules above. The variation contains
the auxiliary field term $\bar{\Psi} \slashed{n}\,
1/(n\cdot \partial)L F\epsilon$ and its conjugate. Since these terms
cannot cancel with any others, we make them vanish by imposing the
constraint (\ref{eq:10}). Four terms then remain in $\delta
S_{\text{new}}$, namely

\begin{equation}
  \label{eq:11}
  \begin{split}
  \delta S_{\text{new}} = \sqrt{2}m^{2}\int\! d^4\!x 
  \biggl[& -\bar{\epsilon} (\slashed{\partial} \bar{Z}) \slashed{n}
    \frac{1}{n\cdot \partial} L\Psi + \bar{\Psi} 
  \slashed{n}  \frac{1}{n\cdot \partial} L (\slashed{\partial}
  Z)\epsilon \\
  &-2 Z \bar{\Psi}R\epsilon -2 \bar{Z} \bar{\epsilon}L\Psi\biggr].
  \end{split}
\end{equation}
The first two terms of (\ref{eq:11}) can be simplified. We temporarily
assume that $\epsilon(x)$ depends on $x^\mu$ to allow determination of
the modified supercurrent. To start the process note that $\slashed{n}
(\slashed{\partial} Z)\epsilon =2(n\cdot \partial Z)\epsilon$ because
$\epsilon$ satisfies (\ref{eq:10}).  The first term of (\ref{eq:11})
then partially integrates to
\begin{equation}
  \label{eq:12}
  \delta S_1 = 2\sqrt{2}m^{2}\int\! d^4\!x
  \left[
    (n\cdot \partial\bar{\epsilon})\bar{Z}\frac{1}{n\cdot\partial}L\Psi
    + \bar{\epsilon}\bar{Z}L\Psi 
  \right],
\end{equation}
in which the last term nicely cancels the last term of (\ref{eq:11}).
In the second term of \eqref{eq:11} we again use (\ref{eq:10}) to
replace $\slashed{n}\slashed{\partial} Z$ with the anti-commutator,
and then write $n\cdot\partial Z \,\epsilon = n\cdot\partial (Z
\,\epsilon) - Z n\cdot\partial \epsilon$. The first part then nicely
cancels the third term in (\ref{eq:11}). This leaves us with
\begin{equation}
  \label{eq:13}
    \delta S_{\text{new}} = 
    2\sqrt{2}m^{2}\int\!d^4\!x \left[(n\cdot \partial\bar{\epsilon})
    \bar{Z}\frac{1}{n\cdot\partial}L\Psi - \bar{\Psi}R
    \frac{1}{n\cdot\partial} (Z n\cdot\partial\epsilon)\right] 
\end{equation}
Since the operator $1/n\cdot\partial$ is anti-Hermitean\footnote{
  $0=\int\! d^4\!x \, n\cdot\partial
  (\frac{1}{n\cdot\partial}\bar{\psi}\frac{1}{n\cdot\partial}\chi) =
  \int\! d^4\!x \,(\frac{1}{n\cdot\partial}\bar{\psi}\chi + \bar{\psi}
  \frac{1}{n\cdot\partial}\chi)$.}, both terms in (\ref{eq:13}) vanish
for constant epsilon,  proving that $S_{\text{new}}$ is invariant
under SUSY variations restricted by the constraint (\ref{eq:10}).

The Noether procedure tells us that the $n\cdot \partial
\bar{\epsilon}$ and $n\cdot \partial \epsilon$ terms in (\ref{eq:13})
modify the supercurrent. The new current includes terms from
$S_{\text{old}}$ and $S_{\text{new}}$:
\begin{equation}
  \label{eq:14}
  \mathcal{J}^\mu=\sqrt{2} \left[L(\slashed{\partial}\bar{Z} +
    F)\gamma^\mu +R(\slashed{\partial}
    Z + \bar{F})\gamma^\mu - 2m^{2} n^\mu(L \bar{Z} + R Z)
    \frac{1}{n\cdot \partial}\right] 
    \Psi 
\end{equation}
Conservation may be checked by contracting with parameters
$\bar{\epsilon}$ which satisfy the constraint (\ref{eq:10}).  Using
the equations of motion it is then not difficult to show that
$\partial_\mu \bar{\epsilon} \mathcal{J}^\mu = 0$. It is also
straightforward to show, using plane wave expansions of the fields,
that $\bar{\epsilon}Q = \int \!d^3\!x \,\bar{\epsilon} \mathcal{J}^0$
generates the conventional SUSY transformations of (\ref{eq:8}).

The SUSY action (\ref{eq:6}) is invariant under spacetime
translations. The conserved energy-momentum tensor can be obtained by
computing the variation of the action with a spacetime dependent
translation vector $a^\mu(x)$.  The stress tensor is the coefficient
of $\partial_\mu a_\nu$:
\begin{multline}
  \label{eq:15}
  T^{\mu\nu}= \bar{\Psi}\gamma^\mu \partial^\nu L\Psi - m^2
  \bar{\Psi}\overleftarrow{\frac{1}{n\cdot
      \partial}} \slashed{n} \frac{1}{n\cdot \partial} n^\mu \partial^\nu
  L\Psi\\
  +\partial^\mu \bar{Z}\partial^\nu Z + \partial^\nu
  \bar{Z}\partial^\mu Z-
  \eta^{\mu\nu}(\partial_\rho\bar{Z}\partial^\rho Z+2m^2 \bar{Z}Z),  
\end{multline}
in which we have dropped terms which vanish by the fermion equation of
motion. Although the fermion terms are not symmetric, this tensor is
conserved on both indices. Thus a symmetric stress tensor is obtained
by the simple device of adding together $\frac{1}{2}(T^{\mu\nu} +
T^{\nu\mu})$. The existence of this symmetric conserved stress tensor
allows us to construct conserved currents corresponding to all six
conventional Lorentz generators:
\begin{equation}
  \label{eq:16}
  M^{\lambda\sigma} = \int\!d^{3}\!x
  \left(
    x^{\lambda} T^{0\sigma} - x^{\sigma} T^{0\lambda}
  \right)
\end{equation}
This indicates that the free theory is actually Lorentz invariant as
discussed in \cite{prep}.  This is as expected; the constraints of
SIM(2) imply Lorentz invariant propagation for all particles. Only in
the presence of interactions will Lorentz violation reveal itself,
through the spin-dependent couplings of the fermion. In the
interacting case the canonical stress tensor will no longer be
conserved on both indices, and a symmetric tensor cannot be
constructed.

Since the SIM(2) modified theory enjoys the usual transformation
rules, conventional superpotential terms and couplings to the gauge
multiplet $A_\mu \equiv T^{a}A_\mu^a,\, \lambda\equiv
T^{a}\lambda^a,\, D^a$ may be introduced. For example, the mass $M$ of
section \ref{sec:sim2-fermions-how} comes from the superpotential.
The general theory is obtained by simply promoting $\partial_\mu
\rightarrow D_\mu = \partial_\mu -i A_\mu$ in the formul\ae{} above
and adding the conventional terms to the action. For completeness this
general SIM(2) SUSY gauge theory is given in the Appendix. The only
new term in the proof of invariance is
\begin{equation}
  \label{eq:17}
  \delta \frac{1}{n\cdot D} = ig \frac{1}{n\cdot D} n^\mu 
  \delta A_\mu \frac{1}{n\cdot D}, 
\end{equation}
but this vanishes because $n^\mu \delta A_\mu = n^\mu
\bar{\epsilon}\gamma_\mu \lambda =0$ by (\ref{eq:10}). 

The SUSY algebra may be deduced by computing the SUSY transform of the
supercurrent. To simplify things we drop the auxiliary field. It is
important to work with conserved components of this current, so we
contract with a spinor parameter constrained by (\ref{eq:10}) and
consider $\bar{\epsilon}\mathcal{J}^\mu$. We then write
\begin{equation}
  \label{eq:18}
 \delta_1 L \bar{\epsilon}_2\mathcal{J}^\mu = \sqrt{2}
 \bar{\epsilon}_2 L\left(\slashed{\partial} 
 \bar{z}\gamma^\mu 
 - 2 m^{2}n^\mu \frac{1}{n\cdot\partial}\right)\delta_1 L\Psi + \cdots.
\end{equation}
All bilinear boson terms are written explicitly, while the ellipses
indicate the fermion terms on which we comment later. After insertion
of $\delta L\Psi$ from (\ref{eq:9}), and some $\gamma$-matrix algebra
in the first term we get
\begin{equation}
  \label{eq:19}
 \slashed{\partial}\bar{z} \gamma^\mu\slashed{\partial} z =
 \tau^\mu_\nu \gamma^\nu -
 \gamma^{\mu\rho\sigma} \partial_\rho\bar{z}\partial_\sigma z, 
\end{equation}
with
\begin{equation}
  \label{eq:20}
 \tau^\mu_\nu = \partial^\mu \bar{z}\partial_\nu z + \partial_\nu
 \bar{z}\partial^\mu z -
 \delta^\mu_\nu \partial_\rho\bar{z}\partial^\rho z. 
\end{equation}
Next we anti-symmetrize in $\epsilon_1 \leftrightarrow \epsilon_2$ to
form the commutator. Using the symmetry properties of Majorana spinor
bilinears, the result simplifies to
\begin{equation}
  \label{eq:21}
 \delta_{[1} L \bar{\epsilon}_{2]}\mathcal{J}^\mu =
 2\bar{\epsilon}_1\left[T^\mu_\nu \gamma^\nu +  2m^2\bar{Z}(\gamma^\mu
   Z - n^\mu \frac{1}{n\cdot \partial} \slashed{\partial}
   Z)\right]\epsilon_2,  
\end{equation}
in which we have added and subtracted the scalar mass term so that
$\tau^\mu_\nu$ is replaced by the stress tensor $T^\mu_\nu$. The first
term is what we expect, but the last term is surely not.  Fortunately
it vanishes due to the spinor condition (\ref{eq:10}) as the following
manipulations show
\begin{equation}
  \label{eq:22}
  \begin{split}
 \bar{\epsilon}_1(\gamma^\mu Z - n^\mu \frac{1}{n\cdot \partial}
 \slashed{\partial} Z)\epsilon_2 &= 
 \bar{\epsilon}_1(\gamma^\mu n\cdot\partial - n^\mu
 \slashed{\partial})\frac{1}{n\cdot \partial} 
  Z)\epsilon_2\\
 &=\bar{\epsilon}_1(\frac{1}{2}
 \gamma^\mu\{\slashed{n},\slashed{\partial}\} - n^\mu 
 \slashed{\partial})\frac{1}{n\cdot \partial}Z)\epsilon_2\\
 &= 0.
  \end{split}
\end{equation}
The final step was achieved by moving the $\slashed{n}$ to the right
or left so that it annihilates the $\epsilon$ spinors.  The spinor
contributions to (\ref{eq:18}) conform to the pattern above and need
not be studied explicitly.

The net result is that the SUSY algebra appears to be conventional,
namely
\begin{equation}
  \label{eq:23}
  \bar{\epsilon_1}\{Q,\bar{Q}\}\epsilon_2
  \,=\,2(\bar{\epsilon}_1\gamma_\mu   \epsilon_2)P^\mu , 
\end{equation}
but it is important that the $\epsilon$ spinors satisfy (\ref{eq:10}).
To incorporate this fact we note that (\ref{eq:10}) can be written as
$\epsilon = \gamma_0\gamma_3\epsilon$, implying that the 4-vector
$\bar{\epsilon}_1\gamma_\mu \epsilon_2$ is proportional to
$n^\mu$. Let $\tilde{n}^\mu$ be any vector whose scalar product with
$n^\mu$ is unity, $\tilde{n}\cdot n =1$.  Then we can rewrite
(\ref{eq:23}) as
\begin{equation}
\label{alg}
 \{\bar{\epsilon_1}Q,\bar{Q}\epsilon_2\}  =2 \Xi_{12}\,n_\mu
 P^\mu
\end{equation}
where $\Xi_{12} = \tilde{n}^\nu (\bar{\epsilon}_1\gamma_\nu
\epsilon_2)$.
Thus only the positive definite combination $n_\mu P^\mu$ 
occurs in the SUSY algebra.

We now consider the determination of the vacuum state in a general
theory of this type. The vacuum must minimize the scalar potential
\begin{equation}
  \label{eq:24}
  V = W'(Z) \bar{W'}(\bar{Z}) + \frac{1}{2} \sum_a
  D^a(Z,\bar{Z})^2 +  \sum 2m^2 \bar{Z}Z,
\end{equation}
where the last sum includes the scalars of all chiral multiplets
with SIM(2) couplings. The vacuum energy must vanish in a
supersymmetric state which requires the 3 conditions
\begin{equation}
  \label{eq:25}
  W'(Z)=0, \qquad  D^a(Z,\bar{Z}) =0 \qquad\text{and}\qquad m^2 Z =0.  
\end{equation}
SUSY is spontaneously broken unless the superpotential gradient and
the $D$-terms vanish at a point where $Z=0$ for all multiplets with
SIM(2) masses.  If SUSY breaking were due only to SIM(2), the vacuum
energy would be of order $V \sim m^2|\langle Z\rangle|^2$ where
$\langle Z\rangle$ is a typical Vev of scalars with SIM(2) masses.
For SIM(2) masses of the order of neutrino masses, $m \approx 10^{-1}$
eV, and similarly sized scalar Vevs, the vacuum energy is near the
experimental value.

\section{SIM(2) and the gauge multiplet}

Consider a free abelian gauge multiplet with physical components
$A_\mu,~\lambda$ and standard SUSY transformation rules
\begin{equation} 
\label{eq:26}
\delta A_\mu = \bar{\epsilon} \gamma_\mu \lambda ~~~
\delta\lambda = \frac{1}{2}
\gamma^{\rho\sigma}F_{\rho\sigma}\epsilon.
\end{equation}
It is not hard to show that the linear equations of motion
\begin{gather}
  \label{eq:26a}
  \left( \slashed{\partial} - m^2 \slashed{n}\frac{1}{n\cdot \partial}
  \right)\lambda =0\\
    \label{eq:26b}
  \left(\partial^\mu - 2m^2 n^\mu \frac{1}{n\cdot \partial}\right)
  F_{\mu\nu}=0
\end{gather}
are covariant under SUSY transformations provided that (\ref{eq:10})
holds. Taking the divergence of the Bianchi identity,
$\partial^\mu \partial_{[\mu} F_{\rho\sigma]}=0,$ and using the
modified Maxwell equation above shows that $(\square -
2m^2)F_{\mu\nu}=0$.

This seems like a promising start for SIM(2) SUSY in the gauge
multiplet, but there are several problems. 
\begin{itemize}
\item The SIM(2) modified Maxwell equation cannot be derived from a
  gauge invariant action.
\item Suppose we add a current source and write
  \begin{equation}
    \label{eq:27}
    \left(\partial^\mu - 2m^2 n^\mu \frac{1}{n\cdot \partial}\right)
    F_{\mu\nu}\,=\, J_\nu.
  \end{equation}
  If we apply $\partial^\nu$ and use (\ref{eq:27}) again, we find
  that the current must satisfy
  \begin{equation}
    \label{eq:28}
    \partial^\nu J_\nu - 2m^2  n^\nu \frac{1}{n\cdot \partial}J_\nu
    \,=\,0,
  \end{equation}
  rather than the usual conservation law.
\item In the non-abelian extension of the construction above, it
appears that there is no nonlinear modification of the equations
of motion which obeys SUSY.
\end{itemize}
 
For these reasons we have not modified the action for gauge
multiplets.

\section{Conclusions}
\label{sec:conclusions}

We have formulated Lorentz violating, but SIM(2) conserving
supersymmetric field theories. The number of supersymmetries is half
that required of a conventional Lorentz invariant theory. Although the
supersymmetry transformations are unmodified, the absence of half the
usual supercharges leads to a modified SUSY algebra involving only
translations along the null direction $n$, rather than the full set of
spacetime translations.

Although the theory is not Lorentz invariant, an old-fashioned
perturbation scheme involving the Hamiltonian may be used to compute
SIM(2) covariant amplitudes \cite{prep}.

SIM(2) theories have novel effects that are forbidden by Lorentz
invariance. The reduced number of supersymmetries in the SIM(2)
context may allow for further interesting physics beyond the standard
model.

Added Note: In \cite{Lindstrom:2006xh} Lindstr\"om and Ro\^cek have
obtained superspace formulations in which both the chiral and
non-abelian gauge multiplet actions are modified by non-local SIM(2)
terms. Their proposal overcomes all difficulties listed in Sec. 4. In
the free limit their gauge multiplet equations of motion agree with
our \eqref{eq:26a}--\eqref{eq:26b}.

\begin{acknowledgments}
  We thank Henriette Elvang, Gary Gibbons, Sheldon Glashow, Matt
  Headrick, Roman Jackiw, and Toby Wiseman for helpful discussions.
  A.~G.~C. was supported in part by the Department of Energy under
  grant no. DE-FG02-01ER-40676.  D.~Z.~F. is supported by NSF grant
  PHY-00-96515 and also by the U.S. Department of Energy under
  cooperative research agreement \#DF-FC02-94ER40818.
\end{acknowledgments}

\appendix*{}

\section{General SIM(2) SUSY gauge theories}

In this appendix, we present the SIM(2) truncation of the general
${\cal N} = 1$ SUSY gauge theory with gauge group G. The theory
contains a chiral matter multiplet $Z^\alpha,L\Psi^\alpha,F^\alpha$
in an arbitrary representation $\mathbf{R}$ of G with Hermitean
generators $(T^a)^\alpha\,_\beta$. The conjugate anti-chiral
multiplet $\bar{Z}_\alpha, R\Psi_\alpha, \bar{F}_\alpha$ is also
required, and these matter multiplets are coupled to the gauge
multiplet $A^a_\mu,\,\lambda^a,\, D^a.$ Representation indices are
suppressed in formulas where this is unambiguous. The component
fields couple through gauge and Yukawa interactions, and an optional
holomorphic gauge invariant superpotential $W(z^\alpha)$.
                                                                                
The covariant derivatives of the various fields are
\begin{subequations}
\begin{align}
  \label{eq:67}
 D_\mu \lambda^a &= \partial_\mu \lambda^a +g f^{abc}A^b_\mu \lambda^c\\
 D_\mu Z &= \partial_\mu Z -igT^a A^a_\mu Z \\
 D_\mu L\Psi &= \partial_\mu L\Psi -igT^a A^a_\mu L\Psi\\
 D_\mu R\Psi &= \partial_\mu R\Psi  +ig R\Psi T^a A^a_\mu\,.
\end{align}
\end{subequations}
                                                                                
The action of the general theory is the sum of several terms
 $S = S_{\text{gauge}} +S_{\text{matter}} + S_{\text{coupling}} +
 S_{F} +S_{\bar{F}}+  S_{\text{new}}$ where
\begin{subequations}
\begin{gather}
  \label{eq:70}
  S_{\text{gauge}} = \int \!d^4\!x
  \left[ -\frac{1}{4}F^{\mu\nu a}F^a_{\mu\nu} +
 \frac{1}{2}\bar{\lambda}^a \gamma^\mu D_\mu \lambda^a +\frac{1}{2}
  D^aD^a\right]\\
  S_{\text{matter}} = \int\! d^4\!x  \left[-D^\mu \bar{Z}D_\mu Z
    +\bar{\Psi}\gamma^\mu L  D_\mu\Psi +\bar{F}F\right]\\
  \label{eq:71}
  S_{\text{coupling}} = g\int\! d^4\!x  \left[-i \sqrt{2}(\bar{\lambda}^a
  \bar{Z}T^a L\Psi -\bar{\Psi}R T^a Z \lambda^a) + D^a \bar{Z}T^a
  Z\right]\\ 
  S_F = -\int\! d^4\!x  \left[F^\alpha W,_\alpha - \frac{1}{2}
    \bar{\Psi}^\alpha L   W,_{\alpha\beta}\Psi^\beta\right]\\
  \label{eq:72}
  S_{\bar{F}} =  -\int\! d^4\!x   \left[\bar{F}_\alpha\bar{W}^\alpha -
    \frac{1}{2}  \bar{\Psi}_\alpha
    R\bar{W},^{\alpha\beta}\Psi_\beta\right]\\ 
  S_{\text{new}} = - m^2\int\! d^4\!x  \left[ \bar{\Psi} \slashed{n}
  \frac{1}{n\cdot \partial}   L\Psi + 2 \bar{Z}Z \right]\,.
\end{gather}
\end{subequations}
The full action is invariant under SUSY transformation rules for the
gauge multiplet
\begin{subequations}
\begin{gather}
  \label{eq:73}
  \delta A^a_\mu = \bar{\epsilon}\gamma_\mu \lambda^a\\
  \delta\lambda^a = [\frac{1}{2} \gamma^{\rho\sigma}F^a_{\rho\sigma}
  +ig D^a]\epsilon\\
  \delta D^a = -\bar{\epsilon} ig \gamma^\mu D_\mu \lambda^a\,.
\end{gather}
\end{subequations}
and for the chiral and anti-chiral multiplets
\begin{subequations}
  \begin{align}
    \label{eq:74}
 \delta Z &= \sqrt{2}\bar{\epsilon}L\Psi &  \delta \bar{Z} &=
 \sqrt{2}\bar{\epsilon}R\Psi\\ 
 \delta L\Psi &= -\sqrt{2} L(\gamma^\mu D_\mu Z + F)\epsilon &  \delta
 R\Psi &= -\sqrt{2} R(\gamma^\mu D_\mu \bar{Z} + \bar{F})\epsilon\\ 
 \delta F &= \sqrt{2} \bar{\epsilon}R(\gamma^\mu D_\mu \Psi
 +ig\lambda^a  T^a Z) &  \delta \bar{F} &= \sqrt{2}
 \bar{\epsilon}L(\gamma^\mu D_\mu \Psi  -ig\lambda^a T^a Z)\,.
  \end{align}
\end{subequations}
These are the conventional transformation rules, but the spinor
$\epsilon$ must satisfy $\slashed{n}\epsilon=0$.
The supercurrent is
\begin{equation}
  \label{eq:75}
  \mathcal{J}^\mu\,=\,\gamma^{\nu\rho}F^a_{\nu\rho}\gamma^\mu \lambda^a+
  \sqrt{2} [L(\slashed{\partial}\bar{Z} + F)\gamma^\mu
  +R(\slashed{\partial} Z + \bar{F})\Psi - 2m^2n^\mu(L \bar{Z} + R Z)
  \frac{1}{n\cdot \partial}] \Psi
\end{equation}
where $F^\alpha=-\bar{W}_{,\alpha},\ \bar{F}^\alpha = - W_{,\alpha}$
is the solution of the auxiliary field equation.  Only the components
obtained by contracting the supercurrent with spinors satisfying the
constraint (\ref{eq:10}) are conserved.


\end{document}